\documentclass[12pt]{article}
 \newcommand{\be}{\begin{equation}}
 \newcommand{\ee}{\end{equation}}
 \newcommand{\bea}{\begin{eqnarray}}
 \newcommand{\eea}{\end{eqnarray}}
 \newcommand{\nn}{\nonumber}

 \date{December 10, 2003}
\title{Complete description of polarization effects in the nonlinear Compton
scattering \\
 {\it II. Linearly polarized laser photons}}

\author{D.Yu.~Ivanov$^{1)}$, G.L.~Kotkin$^{2)}$,
V.G.~Serbo$^{2)}$\\ {\it $^{1)}$Sobolev Institute of Mathematics,
Novosibirsk, 630090, Russia} \\ {\it $^{2)}$Novosibisk State
University, Novosibirsk, 630090, Russia } }

\begin{document}
 \maketitle

 \begin{abstract}

We consider emission of a photon by an electron in the field of a
strong laser wave. Polarization effects in this process are
important for a number of physical problems. We discuss a
probability of this process for linearly polarized laser photons
and for arbitrary polarization of all other particles. We obtain
the complete set of functions which describe such a probability in
a compact form.

 \end{abstract}

\section{Introduction}

Let us consider emission of a photon by an electron in the field
of a strong laser wave. The complete description of polarization
effects in the case of the circularly polarized laser wave was
considered in our paper~\cite{1}. The present paper is a
continuation of Ref.~\cite{1} and we use the same
notations\footnote{Below we shall quote formulas from
paper~\cite{1} by a double numbering, for example, Eq.~(1.21)
means Eq.~(21) from Ref.~\cite{1}.}. Thus, we deal with the
process of nonlinear Compton scattering
 \be
   e(q) +n\,\gamma_L (k) \to e(q') +\gamma (k')\,,
 \label{1}
 \ee
when the electromagnetic laser field is described by 4-potential
 \be
A^\mu(x) = A^\mu\, \cos{(kx)},
 \label{2}
 \ee
where $A^\mu$ is the amplitude of this field. Such a process with
absorbtion of $n=1,\, 2,\, 3,\, 4$ linearly polarized laser
photons was observed in recent experiment at SLAC~\cite{SLAC}.

The method of calculation for such process was developed by
Nikishov and Ritus~\cite{NR-review}. It is based on the exact
solution of the Dirac equation in the external electromagnetic
plane wave. Some particular polarization properties of this
process for the linearly polarized laser photons were considered
in~\cite{NR-review,GR}. In the present paper we give the complete
description of the nonlinear Compton scattering for the case of
linearly polarized laser photons and arbitrary polarization of all
other particles. We follow the method of Nikishov and Ritus in the
form presented in~\cite{BLP} \S 101.

In the next section we briefly describe the kinematics. The cross
section, including polarization of all particles, is obtained in
Sect. 3. In Sect. 4 the polarization of final particles, averaged
over azimuthal angle, is obtained. In Sect. 5 we summarize our
results and compare them with those known in the literature. In
Appendix we give a comparison of the obtained cross section in the
limit of weak laser field with the known results for the linear
Compton scattering.

\section{Kinematics}

All kinematical relations derived in Sect. 2 of Ref.~\cite{1} are
valid for the considered case as well. In particular, the
parameter of nonlinearity is defined via the mean value of squared
4-potential in the same form\footnote{Note, that our definition of
this parameter differs from that used in Refs.~\cite{NR-review} by
additional factor $1/\sqrt{2}$.}
 \be
\xi={e\over mc^2} \sqrt{-\langle A_\mu (x) A^\mu (x) \rangle}=
{e\over mc^2} \sqrt{-\mbox{${1\over 2}$}\, A_\mu A^{\mu}} \,,
 \label{3}
 \ee
where $e$ and $m$ is the electron charge and mass, $c$ is the
velocity of light. Therefore, the amplitude of 4-potential can be
presented in the form
 \be
A^\mu = {\sqrt{2}\, mc^2\over e}\,\xi \;e^\mu_L\,,\;\;\; e_L
e_L=-1\,,
 \label{4}
 \ee
where $e^\mu_L$ is the unit  4-vector describing the polarization
of the laser photons.

In the frame of reference, in which the electron momentum ${\bf
p}$ is anti-parallel to the initial photon momentum ${\bf k}$ (we
call it as a ``collider system''), we direct the $z$-axis along
the initial electron momentum ${\bf p}$. For the discussed problem
it is convenient to choose the $x$-axis along the direction of the
laser linear polarization, i.e. along the vector ${\bf e}_L$ .
Azimuthal angles $\varphi,\; \beta$ and $\beta'$ of vectors ${\bf
k}'$ and electron polarizations $\mbox{\boldmath$\zeta$}$ and
$\mbox{\boldmath$\zeta$}'$ are defined with respect to this
$x$-axis. After that, the unit 4-vector $e_L$ can be presented in
the form
 \be
e_L = e^{(1)}\, \sin{\varphi} -e^{(2)}\, \cos{\varphi}\,,
 \label{5}
 \ee
where the unit 4-vectors $e^{(1)}$ and $e^{(2)}$ are given in
(1.19). This equation determines the quantities $ \sin{\varphi}$
and $ \cos{\varphi}$ in an invariant form, suitable for any frame
of reference. As in paper~\cite{1}, we use the Stokes parameters
$\xi_i$ and $\xi^{\prime}_i$ to describe the polarization of the
initial photon and the detector polarization of the final photon,
respectively. They are defined with respect to the
$x^{\prime}y^{\prime}z^{\prime}$-axes which are fixed to the
scattering plane. The $x'$-axis is the same for both photons and
perpendicular to the scattering plane:
\begin{equation}
x' \;\parallel\; {\bf k} \times {\bf k}';
 \label{6}
\end{equation}
the $y'$-axes are in that plane, in particular,
\begin{equation}
y' \;\parallel\;{\bf k} \times ({\bf k} \times {\bf
k}')=-\omega^2{\bf k}'_{\perp}
 \label{7}
\end{equation}
for the initial photon and
\begin{equation}
y' \;\parallel\; {\bf k}' \times ({\bf k}\times {\bf k}')
  \label{8}
\end{equation}
for the final photon. The azimuthal angle of the linear
polarization of the laser photon in the collider system equals
zero with respect to the $xyz$-axes, and it is $\varphi-(\pi/2)$
with respect to the $x'y'z'$-axes. Therefore, in the considered
case of 100 \% linearly polarized laser beam one has
\begin{equation}
\xi_1=-\sin{2\varphi},\;\;\; \xi_2=0,\;\;\;
\xi_3=-\cos{2\varphi}\,.
 \label{9}
\end{equation}
At small emission angles of the final photon $\theta_\gamma \ll
1$, the final photon moves almost along the direction of the
$z$-axis and ${\bf k} \times {\bf k}'$ azimuth is approximately
equal to $\varphi - (\pi /2)$. Let $\check{\xi}'_j$ be the
detected Stokes parameters for the final photons, fixed to the
$xyz$-axes. They are connected with $\xi'_j$ by the relations
 \be
\xi'_1 \approx -\check{\xi}'_1 \cos{2 \varphi}+ \check{\xi}'_3
\sin{2 \varphi}\,, \; \; \xi'_2 = \check{\xi}' _2 \,,\;\; \xi'_3
\approx -\check{\xi}'_3 \cos{2 \varphi}- \check{\xi}'_1 \sin{2
\varphi}\,.
 \label{10}
 \ee

The polarization properties of the initial and final electrons are
described in the same form as in paper~\cite{1}.

\section{The effective cross section}

As in paper~\cite{1}, we present the effective differential cross
section in the form\footnote{Below we use the system of units in
which $c=1$, $\hbar=1$.}
\begin{equation}
d\sigma (\xi^{\prime}_i,\zeta^{\prime}_j) = {r_e^2\over
4x}\;\sum_n F^{(n)}\;d\Gamma_n\,,\;\;\;d\Gamma_n = \delta
(q+n\,k-q-k')\;{d^3k'\over \omega'}{d^3q'\over q_0'}\,,
 \label{11}
 \end{equation}
where $r_e=\alpha/m$ is the classical electron radius, and
\begin{equation}
F^{(n)}=F_0^{(n)}+\sum ^3_{j=1}\left( F_j^{(n)}\xi '_j\; +
\;G_j^{(n)} \zeta^{\prime}_j\right) + \sum
^3_{i,j=1}H_{ij}^{(n)}\,\zeta^{\prime}_i\,\xi^{\prime}_j \,.
 \label{12}
\end{equation}
We have calculated functions $F_j^{(n)},\, G_j^{(n)}$ and
$H_{ij}^{(n)}$ using the standard technic presented in~\cite{BLP}
\S 101. All the necessary traces have been calculated using the
package MATE\-MA\-TIKA. In the considered case almost all
dependence on the nonlinearity parameter $\xi$ accumulates in
three functions:
 \bea
\tilde{f}_n&=& 4\left[A_1(n,\,a,\,b)\right]^2 -4A_0(n,\,a,\,b)
A_2(n,\,a,\,b) \,,
 \nn\\
\tilde{g}_n &=& {4n^2\over z_n^2}\, \left[A_0(n,\,a,\,b)
\right]^2\,,
 \label{13}
 \\
 \tilde{h}_n &=& {4n\over a}\,A_0(n,\,a,\,b)\,A_1(n,\,a,\,b)\,,
 \nn
 \eea
where functions $A_k(n,\,a,\,b)$ were introduced
in~\cite{NR-review} as follows
 \be
A_k(n,\,a,\,b) =\int\limits_{-\pi}^{\pi} \;\cos^k{\psi}\;\exp
{\left[\,{\rm i}\left(n\psi-a \sin{\psi} + b\sin{2\psi}\right)
\right]}\; {d\psi\over 2\pi}\,.
 \label{14}
 \ee
The arguments of these functions are
 \be
a= e\, \left( {Ap\over kp}- {Ap{^\prime}\over kp'}\right)\,,\;\;
b= {1\over 8}\, e^2 A^2 \,\left( {1\over kp}- {1\over kp'}\right)
 \label{15}
  \ee
or in more convenient variables they are
 \be
a= \sqrt{2}\,\xi\,m\, \left( {e_L p\over kp}- {e_L p'\over
kp'}\right)\,,\;\; b=  {y\over 2(1-y)x}\,\xi^2\,.
 \label{16}
 \ee
In the collider system one has
 \be
 a=- z_n \; \sqrt{2}\, \cos{\varphi}\,,
 \label{17}
 \ee
where
 $$
 z_n= {\xi\over \sqrt{1+\xi^2}}\; n\,s_n
 $$
was defined in (1.41). Among the functions $A_0(n,\,a,\,b)$,
$A_1(n,\,a,\,b)$ and $A_2(n,\,a,\,b)$ there is a useful relation
 \be
(n-2b)\,A_0(n,\,a,\,b)-a\, A_1(n,\,a,\,b)+4b \,A_2(n,\,a,\,b)=0\,.
 \label{18}
  \ee
To find the photon spectrum, one needs also functions (\ref{13})
averaged over the azimuthal angle $\varphi$:
 \be
\langle \tilde{f}_n \rangle =  \int_0^{2\pi} \tilde{f}_n\, {d
\varphi \over 2\pi}\,,\;\; \langle \tilde{g}_n \rangle =
\int_0^{2\pi} \tilde{g}_n\, {d \varphi \over 2\pi}\,.
 \label{19}
 \ee

For small values of $\xi \to 0$ or $y \to 0$, we have
 \bea
a&\propto& \sqrt{y\,\xi^2}\,,\;\;b \propto {y\,\xi^2}\,,
 \label{20}
 \\
A_{0,\,2}(n,\,a,\,b) &\propto& \left( {y\,\xi^2}\,
\right)^{n/2}\,,\;\; A_1(n,\,a,\,b) \propto \left( {y\,\xi^2}\,
\right)^{(n-1)/2}\,,
 \nn
 \\
\tilde{f}_n ,\,\tilde{g}_n ,\, \tilde{h}_n &\propto&
\left({y\,\xi^2}\right)^{n-1}\,,
 \nn
 \eea
in particular, at $\xi=0$ or at  $y=0$
 \be
\tilde{f}_1=\langle \tilde{f}_1 \rangle= \langle \tilde{g}_1
\rangle= \tilde{h}_1=1\,,\;\; \tilde{g}_1=1+\cos2\varphi\,.
 \label{21}
 \ee

The results of our calculations are the following. First, we
define the auxiliary functions
 \bea
&& X_n= \tilde{f}_n-(1+c_n)\left[ (1-\Delta\, r_n)\,\tilde{g}_n
-\tilde{h}_n\,\cos{2\varphi}\right] \, ,
 \nn \\
&& Y_n=(1+c_n)\tilde{g}_n-2\tilde{h}_n \cos^2\!{\varphi} \, ,
 \label{22}
\\
&& V_n= \tilde{f}_n\,\cos{2\varphi} + 2 (1+c_n) \left[(1-\Delta\,
r_n)\,\tilde{g}_n\,-2\tilde{h}_n\, \cos^2\!{\varphi}\,
\right]\,\sin^2\!{\varphi} \,,
 \nn
 \eea
where we use the notation
 $$
 \Delta= {\xi^2 \over 1+\xi^2}\,,
 $$
$c_n$ and $r_n$ is defined in (1.10) and (1.11), respectively.

The item $F_0^{(n)}$, related to the total cross section (1.35),
reads
 \be
F_0^{(n)}={2-2y+y^2\over 1-y}\,\tilde{f}_n- {s_n^2\over 1+\xi^2}\,
\tilde{g}_n \ .
 \label{23}
 \ee

The polarization of the final photons $\xi_j^{(f)}$ is given by
Eq. (1.36) where
  \bea
F_1^{(n)}&=&2\, X_n\, \sin{2\varphi} \,,\;\;\; F_3^{(n)}= -2\,
V_n+\frac{s_n^2}{1+\xi^2}\,\tilde{g}_n \,,
 \label{24}
 \\
F_2^{(n)}&=& {ys_n \over \sqrt{1+\xi^2}} \left[ \, \tilde{h}_n \,
\zeta_1\, \sin{2\varphi}-Y_n\, \zeta_2 \right]+\left( {2-y\over
1-y}\,\tilde{f}_n- {s^2_n\over 1+\xi^2}\,\tilde{g}_n\right) y\,
\zeta_3\,,
 \nn
 \eea

The polarization of the final electrons $\zeta_j^{(f)}$ is given
by Eqs. (1.37), (1.38) with
 \bea
G_1^{(n)}&=&
\left(2\tilde{f}_n-\frac{{s}_n^2}{1+\xi^2}\,\tilde{g}_n\right)\zeta_1+
 {ys_n
 \over (1-y) \sqrt{1+\xi^2}}\, \tilde{h}_n \, \zeta_3 \,\sin{2\varphi}\, ,
 \nn\\
G_2^{(n)}&=&
\left(2\tilde{f}_n-\frac{{s}_n^2}{1+\xi^2}\,\tilde{g}_n\right)\zeta_2-
 {ys_n
 \over (1-y) \sqrt{1+\xi^2}}\, Y_n\zeta_3\,,
 \label{25}\\
G_3^{(n)}&=& - {ys_n \over \sqrt{1+\xi^2}} \left( \, \tilde{h}_n
\, \zeta_1\, \sin{2\varphi}- Y_n \zeta_2 \right)
 \nn
 \\
&& +\left( \frac{2-2y+y^2}{1-y}\tilde{f}_n
-\frac{1-y+y^2}{1-y}\frac{s_n^2}{1+\xi^2} \, \tilde{g}_n
\right)\zeta_3 \, .
 \nn
  \eea

At last, the correlation of the final particles' polarizations are
 \bea
 H_{11}^{(n)}&=&\frac{2-2y+y^2}{1-y}\, X_n \zeta_1\,\sin{2\varphi}
- y\! \left( \frac{2-y}{1-y}\, V_n - {s_n^2
 \over 1+\xi^2}\, \tilde{g}_n \right) \zeta_2 -{ys_n
 \over \sqrt{1+\xi^2}}Y_n   \zeta_3 \, ,
  \nonumber \\
H_{21}^{(n)}&=& \frac{y}{1-y} \left[ (2-y)V_n
-\frac{s_n^2}{1+\xi^2}\, \tilde{g}_n \right]\zeta_1
+\frac{2-2y+y^2}{1-y} X_n \zeta_2\,\sin{2\varphi}
  \nn \\
&& + {ys_n \over \sqrt{1+\xi^2}} \tilde{h}_n \, \zeta_3
\,\sin{2\varphi}\, ,
 \nonumber \\
H_{31}^{(n)}&=&{ys_n  \over (1-y)\sqrt{1+\xi^2}}\left( Y_n \zeta_1
- \tilde{h}_n \,  \zeta_2\, \sin{2\varphi}\right) +2 X_n \zeta_3
\,\sin{2\varphi}\, ,
  \nn \\
H_{12}^{(n)}&=& \frac{y s_n}{(1-y)\sqrt{1+\xi^2}}\, \tilde{h}_n \,
\sin{2\varphi}\,, \ \ \  H_{22}^{(n)}= -\frac{y
s_n}{(1-y)\sqrt{1+\xi^2}}\, Y_n \, ,
 \nn \\
H_{32}^{(n)}&=&\frac{y}{1-y}\left[
(2-y)\tilde{f}_n-\frac{s_n^2}{1+\xi^2}\, \tilde{g}_n \right] \, ,
  \label{26} \\
H_{13}^{(n)}&=& -
\left(\frac{2-2y+y^2}{1-y}V_n-\frac{s_n^2}{1+\xi^2}\,
\tilde{g}_n\right)\zeta_1
-\frac{y(2-y)}{1-y}\,X_n\zeta_2\,\sin{2\varphi}
   \nn \\
&& + \frac{y s_n}{\sqrt{1+\xi^2}}\, \tilde{h}_n\,
\zeta_3\,\sin{2\varphi}\, ,
  \nn \\
H_{23}^{(n)}&=& \frac{y(2-y)}{1-y}\,X_n\zeta_1\,\sin{2\varphi}
-\left(\frac{2-2y+y^2}{1-y}V_n-\frac{1-y+y^2}{1-y}\frac{s_n^2}{1+\xi^2}
\, \tilde{g}_n\right)\zeta_2
  \nn\\
&& +\frac{y s_n}{\sqrt{1+\xi^2}}Y_n \zeta_3\, ,
  \nn \\
H_{33}^{(n)}&=&-\frac{y s_n}{(1-y)\sqrt{1+\xi^2}} \left( \,
\tilde{h}_n\, \zeta_1\,\sin{2\varphi} + Y_n \zeta_2 \right)
-\left( 2\, V_n - \frac{s_n^2}{1+\xi^2}\,
\tilde{g}_n\right)\zeta_3 \, .
  \nn
  \eea

\section{Averaged polarization of the final particles}

In many application it is important to know the averaged over
azimuthal angle $\varphi$ polarization of the final photon and
electron in the collider system. To find it, we can use the same
method as for the linear Compton scattering (see
Refs.~\cite{Khoze}, \cite{GKPS83} and \cite{KPS98}). We substitute
$\xi_j$, $\zeta_j$, $\xi'_j$ and $\zeta'_j$ from Eqs. (\ref{9}),
(1.51), (\ref{10}) and (1.56), respectively, into Eq. (\ref{12})
and obtain after integration over azimuthal angle $\varphi$:
 \bea
&&{d\sigma_n (\xi^{\prime}_i,\zeta^{\prime}_j)\over dy} \approx
{\pi r_e^2\over 2x}\;\left( \langle F_0^{(n)} \rangle + \langle
\Phi_1^{(n)} \rangle \,\check{\xi}'_1 + \langle F_2^{(n)}
\rangle\, \check{\xi}'_2+ \langle\Phi_3^{(n)} \rangle\,
\check{\xi}'_3 + \right.
 \label{26a}
 \\
&&\left.+ G_\perp^{(n)} \mbox{\boldmath$\zeta$}_\perp
\mbox{\boldmath$\zeta$}'_\perp + G_\parallel^{(n)} \zeta_3
\zeta_3' + \sum^3_{i,j=1} \langle
H_{ij}^{(n)}\,\zeta^{\prime}_i\,\xi^{\prime}_j \rangle \right)\,,
 \nn
 \eea
where
 \bea
\langle F_0^{(n)} \rangle&=& \left({1\over
1-y}+1-y\right)\,\langle \tilde{f}_n \rangle- {s_n^2\over
1+\xi^2}\,\langle \tilde{g}_n \rangle\,,
 \nn
 \\
\langle \Phi_1^{(n)} \rangle&=&\langle -F_{1}^{(n)}
\,\cos{2\varphi}-F_{3}^{(n)}\,\sin{2\varphi} \rangle =0\,,
 \nn
 \\
\langle F_2^{(n)} \rangle &= &\left( {2-y\over
1-y}\,\langle\tilde{f}_n \rangle- {s^2_n\over
1+\xi^2}\,\langle\tilde{g}_n\rangle \right) y\, \zeta_3\,,
 \nn
 \\
\langle \Phi_3^{(n)} \rangle &=& \langle
F_{1}^{(n)}\,\sin{2\varphi}-F_{3}^{(n)}\,\cos{2\varphi} \rangle\,,
 \nn
 \\
G_\perp^{(n)}&=&
2\langle\tilde{f}_n\rangle-\frac{{s}_n^2}{1+\xi^2}\,
\langle\tilde{g}_n\rangle\,,
 \nn
 \\
G_\parallel^{(n)} &=&  \frac{2-2y+y^2}{1-y}\langle\tilde{f}_n
\rangle -\frac{1-y+y^2}{1-y}\frac{s_n^2}{1+\xi^2} \,
\langle\tilde{g}_n \rangle  \,,
 \nn
 \eea
As a result, the averaged Stokes parameters of the final photon
are
  \be
\langle \check\xi_{1(n)}^{(f)} \rangle \approx 0\,,\;\;
 \langle \check\xi_{2(n)}^{(f)} \rangle \approx
 {\langle F_2^{(n)} \rangle\over \langle F_{0}^{(n)}
 \rangle}\,,\;\;
\langle \check\xi_{3(n)}^{(f)} \rangle \approx {\langle
\Phi_{3}^{(n)} \rangle \over \langle F_{0}^{(n)} \rangle}\,,
  \ee
and the averaged polarization of the final electron is
 \be
 \langle \mbox{\boldmath$\zeta$}^{(f)}_{\perp (n)} \rangle  \approx
 {G_\perp^{(n)} \over \langle F_{0}^{(n)} \rangle}\,
 \mbox{\boldmath$\zeta$}_\perp \,,\;\;
 \langle \zeta_{3(n)}^{(f)} \rangle \approx  {G_\parallel^{(n)}\over \langle
F_{0}^{(n)}\rangle }\, \zeta_3\,.
 \ee

Note that averaged Stokes parameters of the final photon do not
depend on $\mbox{\boldmath$\zeta$}_\perp$ and that averaged
polarization vector of the final electron is not equal zero only
if $\mbox{\boldmath$\zeta$} \neq 0$.  These properties are similar
to those in the linear Compton scattering.

\section{Summary and comparison with other papers}

Our main result is given by Eqs. (\ref{23})--(\ref{26}) which
present 16 functions $F_0,\; F_j,\; G_j$ and $H_{ij}$ with $i,\,j
= 1\div 3$. They describe completely all polarization properties
of the nonlinear Compton scattering in a rather compact form.

In the literature we found 4 functions which can be compared with
ours $F_0, F_1,\, F_2,\, F_3$. They enter the total cross section
(1.35), (1.54), the differential cross sections (1.53), (1.54) and
the Stokes parameters of the final photons (1.36). The function
$F_0$ was calculated in~\cite{NR-review}, the functions $F_j$ were
obtained in~\cite{GR}. Our results (\ref{23}), (\ref{24}) coincide
with the above mentioned ones.

The polarization of the final electrons is described by functions
$G_j$ (\ref{25}). They enter the polarization vector
$\mbox{\boldmath$\zeta$}^{(f)}$ given by exact (1.38), (1.61) and
approximate (1.67) equations.  The correlation of the final
particles' polarizations are described by functions $H_{ij}$ given
in Eqs. (\ref{26}). We did not find in the literature any on these
functions.

At small $\xi^2$ all harmonics with $n > 1$ disappear due to
properties (\ref{20}) and (\ref{21}),
 \be
d\sigma_n(\xi^{\prime}_i,\zeta^{\prime}_j) \propto \xi^{2(n-1)}
\;\; \mbox{ at } \;\; \xi^2\to 0\,.
 \label{27}
 \ee
We checked that in this limit our expression for $d\sigma
(\xi^{\prime}_i,\zeta^{\prime}_j)$ coincides with the result known
for the linear Compton effect, see Appendix.

\section*{Acknowledgements}

We are grateful to I.~Ginzburg, M.~Galynskii, A.~Milshtein,
S.~Polityko and V.~Telnov for useful discussions. This work is
partly supported by INTAS (code 00-00679) and RFBR (code
02-02-17884); D.Yu.I. acknowledges the support of Alexander von
Humboldt Foundation.

\section*{Appendix: Limit of the weak laser field}

At $\xi^2\to 0$, the cross section (\ref{11}) has the form
\begin{equation}
d\sigma (\xi^{\prime}_i,\zeta^{\prime}_j) = {r_e^2\over 4x}\;F
\;d\Gamma;\;\;\;d\Gamma = \delta (p+k-p-k')\;{d^3k'\over
\omega'}{d^3p'\over E'}\,,
 \label{28}
 \end{equation}
where
\begin{equation}
F=F_0+\sum ^3_{j=1}\left( F_j\xi'_j\; + \;G_j
\zeta^{\prime}_j\right) +
\sum^3_{i,j=1}H_{ij}\,\zeta^{\prime}_i\,\xi^{\prime}_j \,.
 \label{29}
\end{equation}
To compare it with the cross section for the linear Compton
scattering, we should take into account that the Stokes parameters
of the initial photon have values (\ref{9}) and that our
invariants $c_1$, $s_1$, $r_1$ and auxiliary functions (\ref{22})
transforms at $\xi^2=0$ to
 \bea
c_1&\to& c=1- 2r\,,\;\; s_1\to s =2\sqrt{r(1-r)}\,,\;\; r_1\to
r={y\over x(1-y)}\,,
  \label{30}
  \\
X_1 &\to& -c\,,\;\; Y_1 \to c(1-\xi_3)\,,\;\; V_1 \to -\xi_3\,.
 \nn
 \eea
Our functions
 \bea
F_0&=&{1\over 1-y}+1-y-s^2(1-\xi_3)\,,\;\; F_1=2c \xi_1\,,\;\;
F_3= s^2+(1+c^2)\,\xi_3\,,
 \label{31}
 \\
F_2&=&- {ys c}\,\zeta_2+
 y\left( {1\over 1-y}\,+ c^2\right) \zeta_3
 -ys\,\zeta_1 \xi_1+
 ys\left(c\zeta_2+s\zeta_3 \right)\, \xi_3
 \nn
 \eea
coincide with those in~\cite{GKPS83}. Our functions
 \bea
G_1&=&(1+c^2+s^2\xi_3)\zeta_1-{y\,s\over 1-y}\xi_1 \zeta_3\,,
 \nn
  \\
G_2 &=& (1+ c^2+ s^2 \xi_3 )\; \zeta_2 - {ysc \over 1-y} \, (1-
\xi_3 ) \;\zeta_3\,,
 \label{32}
 \\
G_3&=& ys \xi_1 \zeta_1 + ysc (1 -\xi_3)\; \zeta_2+\left[1+ \left(
{1\over 1-y} -y \right)\,(c^2 +s^2 \xi_3 ) \right] \; \zeta_3.
 \nn
 \eea
coincide with functions $\Phi_j$ given by Eqs. (31)
in~\cite{KPS98}. At last, we check that our functions
 \bea
H_{11}&=&{2-2y+y^2\over 1-y}\,c\,\zeta_1\,\xi_1 +y\left[ {2-y\over
1-y}\,\xi_3 +s^2(1-\xi_3)\right]\,\zeta_2
-ysc(1-\xi_3)\,\zeta_3\,,
 \nn\\
 H_{21}&=&-\frac{y}{1-y}\left[
s^2+(1-y+c^2)\xi_3\right]\zeta_1+ {2-2y+y^2\over
1-y}\,c\xi_1\zeta_2 -ys\xi_1\zeta_3\, ,
 \nonumber \\
H_{31}&=& \frac{y c s}{1-y}\,(1-\xi_3) \zeta_1+\frac{y
s}{1-y}\,\xi_1 \zeta_2 +2c\xi_1\zeta_3 \, ,\;\;
 H_{12}= -{ys\over 1-y}\, \xi_1 \, ,
 \label{33} \\
H_{22}&=&-\frac{y c s}{1-y} \,(1-\xi_3) \,,\;\;
H_{32}=\frac{y}{1-y} \left[2-y-s^2(1-\xi_3)\right] \, ,
 \nonumber\\
H_{13}&= &s^2\,\zeta_1+ \left(1+c^2 +\frac{y^2}{1-y}\right)\,
\xi_3 \zeta_1 -yc\,{2-y\over 1-y}\,\xi_1\, \zeta_2-
ys\,\xi_1\zeta_3\,,
 \nn\\
H_{23}&=&y\frac{2-y}{1-y}\,c\,\xi_1\, \zeta_1+
\frac{1-y+y^2}{1-y}\,s^2\,\zeta_2 +\left(1+c^2
+\frac{y^2c^2}{1-y}\right)\,\xi_3\zeta_2+ysc(1-\xi_3)\,\zeta_3 \,,
 \nonumber \\
H_{33}&=&\frac{y s}{1-y}\,\xi_1\, \zeta_1 -
\frac{ysc}{1-y}\,(1-\xi_3)\,\zeta_2+ s^2\,\zeta_3+(1+c^2)
\,\xi_3\,\zeta_3
 \nn
 \eea
coincide with the corresponding functions from \cite{Grozin}. At
such a comparison, one needs to take into account that the set of
unit 4-vectors, used in our paper (see Eqs. (1.19), (1.23)), and
one used in paper \cite{Grozin} are different. The relations
between our notations and the notations used in~\cite{Grozin} are
given by Eqs. (1.75).

\end{document}